\documentclass[submission,Phys]{SciPost}
\usepackage{amsmath,cases}
\usepackage[utf8]{inputenc}
\DeclareUnicodeCharacter{2215}{/}
\usepackage[T1]{fontenc}
\usepackage{lmodern}
\usepackage{graphicx}
\usepackage{amssymb}
\usepackage{gensymb}
\usepackage{bm}
\usepackage{tabularx}
\usepackage{mathtools}
\usepackage{microtype}
\usepackage{siunitx}
\usepackage[version=4]{mhchem}
\usepackage{bbm}
\usepackage{braket}
\usepackage{mathtools}
\usepackage[normalem]{ulem}
\usepackage{layouts}
\usepackage{float}
\usepackage{textgreek}
\usepackage{orcidlink}

\usepackage{datatool, filecontents}
\DTLsetseparator{,}

\hypersetup{colorlinks=true, linkcolor=blue, citecolor=blue, urlcolor=blue}

\newcounter{CommentNumber}
\renewcommand{\paragraph}[1]{\stepcounter{CommentNumber}\belowpdfbookmark{#1}{\arabic{CommentNumber}}}

\begin{document}

\begin{center}{\Large \textbf{
    Gate-defined Kondo lattices with valley-helical quantum dot arrays
}}\end{center}

\begin{center}
    Antonio L.~R.~Manesco\textsuperscript{1, $\dagger$}\orcidlink{0000-0001-7667-6283}
\end{center}
\begin{center}
    {\bf 1} Kavli Institute of Nanoscience, Delft University of Technology, Delft 2600 GA, The Netherlands
    \\
    \textsuperscript{$\dagger$}am@antoniomanesco.org
\end{center}
\begin{center}
    May 30, 2025
\end{center}

\section*{Abstract}

Kondo physics and heavy-fermion behavior has been predicted and observed in moiré materials.
The electric tunability of moiré materials allows an \emph{in-situ} study of Kondo lattices' phase diagrams, which is not possible with their intermetallic counterparts.
However, moiré platforms rely on twisting, which introduces twisting angle disorder and undesired buckling.
Here we propose device layouts for one- and two-dimensional gate-defined superlattices in Bernal bilayer graphene where localized states couple to dispersive valley-helical modes.
We show that, under electronic interactions, these superlattices are described by an electrically-tunable Kondo-Heisenberg model.

\section{Introduction}

\paragraph{It is hard to study the phase diagram of intermetallic Kondo lattices.}

Kondo lattices are constituted by an array of magnetic moments coupled to a dispersive electron gas~\cite{mott1974rare,doniach1977kondo,coleman2015heavy}.
Historically, these lattices were first realized in intermetalic compounds, where localized $d$ or $f$ orbitals couple to dispersive bands leading to heavy fermion behavior~\cite{PhysRevLett.35.1779}.
The large charge density of these compounds screens electric fields, forbiding electrostatic tunability of the electron density.
Therefore an exploration of the phase diagram requires the growth of several samples~\cite{grosche2001superconductivity,Custers_2003,friedemann2009detaching,park2006hidden,Schr_der_2000,shishido2005drastic,gegenwart2007multiple,Friedemann_2010}.

\paragraph{Moiré Kondo lattices are electrically-tunable platforms.}

In the recent years, moiré materials became standard platforms to study correlated matter~\cite{Cao2018,cao2018unconventional,andrei2021marvels}.
The quench of the kinetic energy caused by the moiré potential enhances the effects of electronic interactions, triggering correlated phases~\cite{bistritzer2011moire,andrei2020graphene}.
In the past few years, moiré Kondo lattices have been both theoretically proposed and probed experimentally~\cite{kumar2022gate,ramires2021emulating,dalal2021orbitally,guerci2023chiral,vavno2021artificial}.
In contrast to their intermetallic counterparts, the electric tunability of moiré Kondo lattices allow \emph{in situ} control of the phase diagram.

\paragraph{Gate-defined devices are an alternative to overcome limited quality of moiré materials.}

Despite their versatility and tunability, experiments in moiré materials lack quantitative reproducibility~\cite{lau2022reproducibility}.
The mechanical manipulation required to fabricate moiré devices limit their quality.
Thus, their electronic structures vary from one sample to the other.
These fabrication issues also limit the scalability of moiré devices.
Gate-defined superlattices overcome these limitations.
The recent progress on patterning graphite gates allows manipulation of the superlattice potential without mechanical manipulation of the active material layer~\cite{Cohen2023}.
Moreover, unlike moiré materials, gate-defined superlattices are not constrained by the crystal lattice of the material.

\paragraph{Here we propose a platform for gate-defined Kondo lattices.}

Here we propose how gate-defined Bernal bilayer graphene superlattices as a platform for electrically controllable Kondo systems.
Our proposal is inspired by recent experiments showing valley filtering via valley-helical channels~\cite{Li2016,Kang2018,Li2018,PhysRevApplied.11.044033,chen2020gate,Huang_2024,davydov2024easy}.
We propose a gate geometry that creates a series of quantum dots coupled by these helical channels.
We build a low-energy model for this device and show that, under electronic interactions, this Hamiltonian maps to a Kondo lattice.
Finally, we extend this device to two-dimensional geometries.
Our proposal is similar to Kondo lattices in twisted trilayer graphene~\cite{ramires2021emulating}, but in a fully electrically-defined setup.

\section{Valley-helical quantum dot arrays}
\label{sec:helical}

\paragraph{We consider a zigzag gate geometry.}

\begin{figure}
    \includegraphics[width=\textwidth]{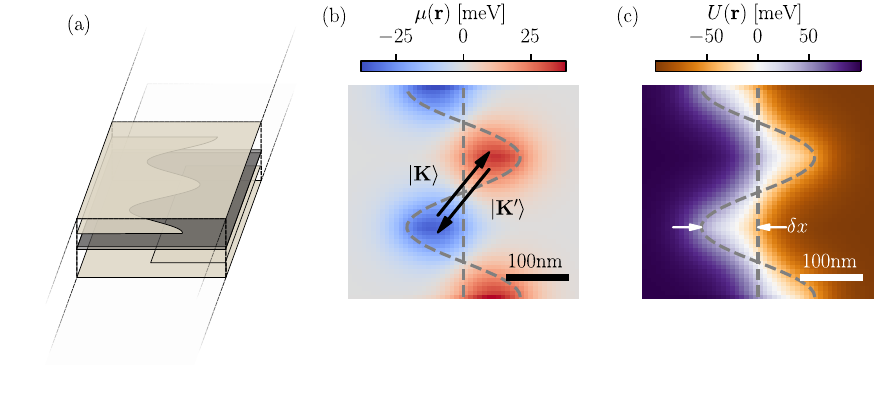}
    \caption{
        (a) Proposed device layout: a Bernal bilayer graphene sheet (black layers) with two top and two bottom gates.
        The outermost gates cover the full device, while the middle gate layers are patterned and screen the outer gates.
        The resulting chemical potential (a) and layer imbalance (b) result in a series of charge puddles with alternating polarities surronded by a displacement field that flips direction from one side to the other of the device.
        The electrostatic profile creates a series of quantum dots coupled by valley-helical point contacts.
    }
    \label{fig:device}
\end{figure}

We present a scheme of the device in Fig.~\ref{fig:device}(a).
Our layout consists of Bernal bilayer graphene with two bottom and two top gates, creating the chemical potential and displacement field landscapes shown in Fig.~\ref{fig:device}~(b, c).
The regions with nonzero displacement field in Fig.~\ref{fig:device}~(b) result in a series of puddles with alternating polarity.
These puddles are surrounded by regions with nonzero displacement field shown in Fig.~\ref{fig:device}(c).
Because displacement fields gap the electronic structure of bilayer graphene (we review effects of displacement fields in Appendix \ref{sec:displacement}), the electrostatic profile results in a series of quantum dots with alternating polarity~\cite{McCann_2006,PhysRevLett.99.216802,Min_2007,Oostinga2007,Zhang2009,PhysRevB.80.165406,Mak2009,Weitz2010}.
The connections between these quantum dots are point contacts constrained by regions with opposite displacement field.
Since the valley Chern number flips sign with the displacement field (see Appendix \ref{sec:displacement}), these point contacts are valley-helical~\cite{Martin2008,Jung2011, Li2016}, as illustrated by the arrows in Fig.~\ref{fig:device}(b).

\paragraph{We thus observe dot-like states weakly coupled to helical states.}

\begin{figure}
    \includegraphics[width=\textwidth]{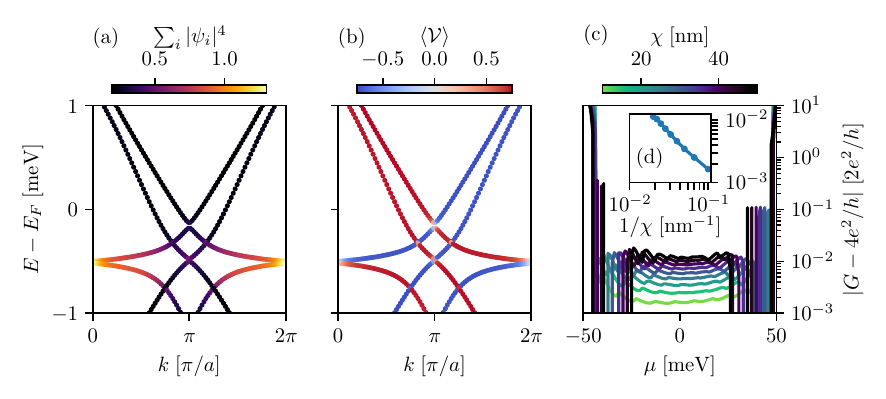}
    \caption{
        Bandstructure of a unit cell of the device shown in Fig.~\ref{fig:device}.
        In panel (a) the bandstructure is colored according to its inverse participation ratio (IPR), while in (b) the colors indicate valley polarization.
        We observe that localized states (with large IPR) couple to highly-dispersive valley-helical channels preserving valley number.
        (c) Conductance accross an unit cell of the device.
        We observe that $G\approx4e^2/h$ indicating that the transport is dominated by the valley-helical channels.
        The inset shows the scaling of $|G - 4e^2/h|$ at $\mu=0$ as a function of $1/\xi$.
        The scaling shown in the inset occurs due to the suppression of the overlap between dot levels.
    }
    \label{fig:helicity}
\end{figure}

In Fig.~\ref{fig:helicity} we show the low-energy bandstructure of a repeating unit cell of this device.
We performed the tight-binding atomistic simulations using Kwant~\cite{Groth2014} and MeanFi~\cite{meanfi} and provide all the details in the Appendix \ref{sec:simulation_details}.
In Fig.~\ref{fig:helicity} (a) the bands are colored according to their inverse participation ratio (IPR).
We observe that the presence of nearly flat bands with high IPR, which correspond to states localized in the charge puddles shown in Fig.~\ref{fig:device} (b).
These localized states are weakly coupled to highly dispersive states.
In Fig.~\ref{fig:helicity} (b) we show the valley-resolved bandstructure, where we can see that these dispersive states are valley helical.
These states are weakly coupled to the dot levels with a strength $\Delta \sim \hbar v_F / \chi \sim 0.1 - 1\unit{meV}$, where $\chi$ is the screening length of the electrostatic profile.
These estimation of $\Delta$ agrees with the observed anticrossings in the bandstructure.
Because intervalley coupling depends on the small parameter $1/K\chi$, where $K$ is the valley momentum, we observe that valley anticrossings are negligible.
This observation is consistent with experimental results showing valley convervation in gate-defined nanostructures~\cite{Eich2018,Moller2023_competing_dots,davydov2024easy,Banszerus2018_ehdots,Ingla2023,Gold2021,InglaAyns2023,Banszerus2021_tunable_dots,Banszerus2020_sogap_qpc,Huang2024}, and justifies our choice to neglect intervalley scattering in the construction of a minimal model.
These experiments also suggest that the disorder in the state-of-the-art devices is small or sufficiently screened by the gates.

\paragraph{The helical channels bottleneck the transmission.}

The dot levels at different charge puddles only couple through the valley helical states.
We demonstrate the lack of direct interdot coupling computing the conductance accross an unit cells of this device.
Since a perfectly helical point contact has a conductance $G=4e^2/h$, we compute the difference $|G - 4e^2/h|$, shown in Fig.~\ref{fig:helicity}(c).
The saturation of conductance at $4e^2/h$ indicates that only two helical valley states per spin channel propagate through the device.
Moreover, in the inset of Fig.~\ref{fig:helicity}(c) we show the suppression of the conductance as a function of the screening length.
Tunneling between the dots is suppressed by the displacement field.
Thus, soft walls caused by large $\chi$ increase the overlap between the dot states, increasing tunneling of non-helical channels.
Due to possible challenges on fabricating the multi-gated layout shown in Fig.~\ref{fig:device}(a), we suggest that an alternative layout with a single layer of split gates similar to a recent experimental realization is possible~\cite{davydov2024easy}.
We show simulations of the split-gate geometry in Appendix~\ref{app:split_gates}, and emphasize that due to the large size of the charge puddles we do not expect the required charging energies to observe Kondo pheomena discussed in Sec.~\ref{sec:kondo}.

\paragraph{We write an effective model for dot coupled to helical states.}

\begin{figure}
    \includegraphics[width=\textwidth]{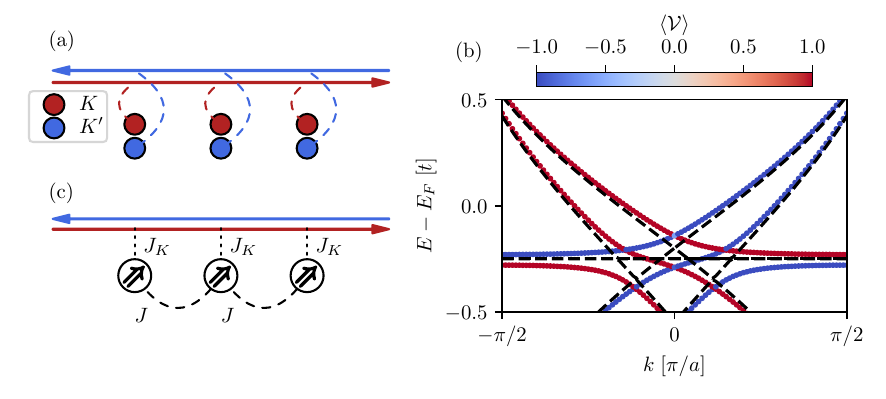}
    \caption{
        (a) Scheme of non-interacting model described by the Hamiltonian in Eq.~\ref{eq:h_eff}.
        Helical valley states couple to dot levels preserving valley number.
        (b) Bandstructure from a lattice model of Eq.~\ref{eq:h_eff} using the tangent fermions approach.
        We recover the features of the bandstructure shown in Fig.~\ref{fig:helicity} (b).
        Dashed lines show the dispersion when $\Delta=0$.
        (c) Scheme of low-energy interacting model.
        Due to a charging energy $V \gg \Gamma, \Delta$ we obtain a gate-tunable Kondo-Heisenberg model.
    }
    \label{fig:effective}
\end{figure}

From the analysis above, we can construct a low-energy theory of helical states coupled to dot levels.
We split the effective Hamiltonian in three terms
\begin{equation}
    \label{eq:h_eff}
    H(k) = H_{\text{helical}}(k) + H_{\text{dot}} + H_{\Delta}(k)~.
\end{equation}
Here, $H_{\text{helical}}$ is the Hamiltonian of the helical states that are perfectly transmited through the point contacs.
Each valley has two helical channels that we distinguish with the index $\rho=\pm$.
The Hamiltonian of the helical states is 
\begin{equation}
    H_{\text{helical}}(k) = \sum_{\tau, \sigma, n} \tau (\rho^{\delta} + v^{\rho} k) c_{\tau \sigma \rho}^{\dagger}(k)c_{\tau \sigma \rho}(k)
\end{equation}
where $v^{\rho}$ is the Fermi velocity of the helical mode $\rho$, $\delta^{\rho}$ is the offset of each helical mode at $k=0$, $\tau = \pm$ is the valley index, $\sigma$ is the spin index, and $k$ is the momentum.
The Hamiltonian of a quantum dot level at the site $i$ is
\begin{equation}
    H_{\text{dot}} = \epsilon \sum_{\tau \sigma i} d_{\tau \sigma i}^{\dagger} d_{\tau, \sigma i}
\end{equation}
where $\epsilon$ is the energy of the dot levels.
For simplicity, we ignore higher energy quantum dot levels and assume the quantum dots in all the sites $i$ are degenerate.
Finally, the valley-preserving hopping between helical and dot states is
\begin{equation}
    H_{\Delta}(k) = \Delta \sum_{i, \tau, \sigma, \rho} d_{\tau \sigma i}^{\dagger}c_{\tau \sigma \rho}(k) + h.c.~.
\end{equation}
A scheme of this effective model is shown in Fig.~\ref{fig:effective} (a).
We implement this model on a lattice using the tangent fermions approach~\cite{PhysRevD.26.468,Beenakker_2023} (see Appendix \ref{sec:tangent}), and recover a dispersion similar to the one shown in Fig.~\ref{fig:helicity} (b).

\section{Effective Kondo lattice model}
\label{sec:kondo}

\paragraph{Interactions lead to an effective Kondo problem.}

We now consider effects of electronic interactions.
Interactions in the helical states only renormalize $v^{\rho}$ and are absorbed by $H_{\text{helical}}(k)$~\cite{Gonz_lez_1994,Kotov_2012,Elias_2011,Mishchenko_2007,Sheehy_2007,Stauber_2017}.
Thus, we are left with the Coulomb interaction between dot levels
\begin{equation}
    H_{\text{dot-dot}} = V \sum_{\tau, \tau^{\prime} \sigma, \sigma^{\prime}, i} n_{\tau\sigma i}n_{\tau^{\prime}, \sigma^{\prime} i}~,
\end{equation}
where $V$ is the charging energy of the quantum dots and $n_{\tau, \sigma i} = d_{\tau, \sigma i}^{\dagger} d_{\tau, \sigma i}$.
We estimate that the charging energy is similar to other graphene quantum dots of comparable sizes, $V\sim 5-10\unit{meV}$~\cite{Güttinger_2012}.
Perturbative expansion over the small parameter $\Delta/V$ reduces the problem to an SU(4) Kondo model~\cite{Choi_2005,ramires2021emulating}
\begin{equation}
    H(k) \approx H_{\text{helical}}(k) + \sum_{i,\rho, \rho^{\prime}, k^{\prime}} J_K^{\rho \rho^{\prime}, kk^{\prime}} \mathbf{S}_i \cdot (\psi_{\rho^{\prime} k^{\prime}}^\dagger \boldsymbol{\gamma} \psi_{\rho k})~,
\end{equation}
where $J_K \sim \Delta^2/V$, $\mathbf{S}_i = \Psi_i^{\dagger} \boldsymbol{\gamma} \Psi_i$ is the $SU(4)$ spin operator of the quantum dot at the site $i$ with $\Psi_i = (d_{+\uparrow i}, d_{+\downarrow i}, d_{-\uparrow i}, d_{-\downarrow i})^T$, $\boldsymbol{\gamma}$ is a vector where the entries are the generators of $SU(4)$, and $\psi_{\rho k} = (c_{+\uparrow\rho}(k), c_{+\downarrow\rho}(k), c_{-\uparrow\rho}(k), c_{-\downarrow\rho}(k))^T$.
Note that Kondo temperature of an $SU(N)$ lattice increases exponentially with $N$~\cite{Choi_2005,PhysRevB.90.121406,PhysRevB.28.5255,Ramires_2016,PhysRevLett.57.877,PhysRevB.35.3394,PhysRevB.38.316}.

\paragraph{The interdot tunneling gives an additional exchange term that competes with $J_K$.}

In the Hamiltonian of Eq.~\ref{eq:h_eff}, we neglected the interdot direct tunneling $\Gamma$ since it is exponentially suppressed by the gap caused by the displacement field, $\Gamma \sim \exp(-U)$.
However, the in-gap transmission shown in Fig.~\ref{fig:helicity} (c) suggests a small yet finite interdot tunneling rate $\Gamma$.
The scaling in Fig.~\ref{fig:helicity} (d) shows that the hopping probability across a unit cell of the device is $T \sim \chi$.
Because an electron must hop twice to cross a unit cell and the transmission probability across each tunnel barrier is $\sim \Gamma^2$, we can estimate the interdot tunneling rate as $\Gamma \sim T^{1/4} \sim \chi^{1/4}$.
Within the tunneling regime, $\Gamma/V \ll 1$, and we can again take the lowest order in perturbation theory, leading to an additional Heisenberg-like exchange coupling~\cite{PhysRevB.109.245401,ramires2021emulating}
\begin{equation}
    H_{\text{exchange}} = J\sum_{\langle i, j\rangle} \mathbf{S}_i \cdot \mathbf{S}_j~,
\end{equation}
with $J\sim \Gamma^2/V$.
Moreover, we discussed in Sec.~\ref{sec:helical} that $\Delta \sim \chi^{-1}$ and is weakly dependent on $U$.
Therefore, since the ratio $\Delta / \Gamma$ can be tunned with electrostatic engineering, this device is described by an electrically-tunable Kondo-Heisenberg Hamiltonian.
An estimation of realistic parameter ranges requires electrostatic simulations that are beyond the scope of this work.

\begin{figure}[!t]
    \includegraphics[width=\textwidth]{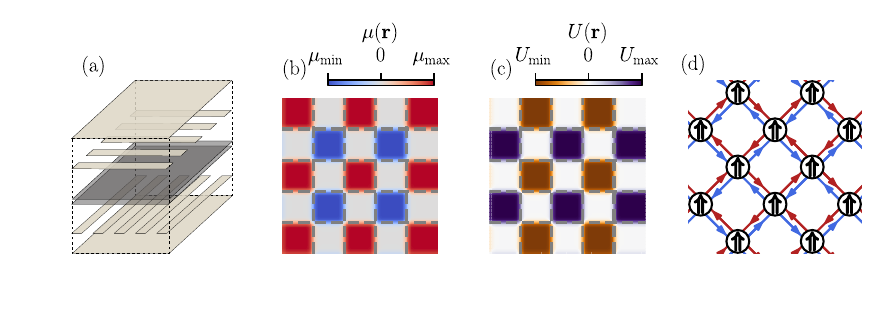}
    \caption{
        (a) Proposed device layout for a two-dimensional geometry.
        This layout results in the chemical potential landscape shown in (a) and layer imbalance shown in (b).
        The checkerboard pattern of charge puddles are connected by valley-helical point.
        (d) The low-energy description of this device consists of a series of localized moments in the quantum dots connnected by two copies of Chaker-Coddington networks with valley-dependent chirality.
    }
    \label{fig:2d_model}
\end{figure}

\paragraph{It is possible to extend this to two-dimensional geometries.}

While we provide a gate layout for an one-dimensional system, this construction can be extended to two-dimensional lattices.
In Fig.~\ref{fig:2d_model}(a) we show a layout proposal for a two-dimensional lattice of quantum dots connected by valley-helical channels.
This gate configuration creates a checkerboard landscape for the chemical potential and displacement field profiles, shown in Fig.~\ref{fig:2d_model} (b, c).
The connections between the quantum dots results in two copies of a Chalker-Coddington network model~\cite{chalker1988percolation} with valley-dependent chirality, \emph{i.e.} a valley-helical network.
Similar networks were predicted to existed in both twisted multilayers and strained graphene superlattices~\cite{ramires2021emulating,PhysRevB.88.121408,PhysRevB.88.121408,Mahmud_2023,PhysRevB.102.245427,Manesco_2021,PhysRevB.98.035404}.
Since we expect similar energy scales similar to the one-dimensional system in this two-dimensional lattice, a similar analysis results again in a gate-controllable Kondo-Heisenberg model.
Unlike the one-dimensional geometry shown in Fig.~\ref{fig:device}(a), this two-dimensional geometry is roubust against relative twists between the two gate layers.

\section{Conclusion}

We presented a strategy to obtain gate-defined graphene superlattices where quantum dots are coupled by valley helical quantum point contacts.
We showed that the low-energy description of these superlattices maps to a electrically-tunable Kondo-Heisenberg Hamiltonian.
We proposed a gate layouts for both one- and two-dimensional superlattices.
Furthermore, we showed that the helical modes in two-dimensional superlattices form a gate-defined valley-helical network model.
Finally, we introduced a lattice description using the tangent fermions approach that can be used to simulate these superlattices with a reduced computational cost.

\section*{Acknowledgements}

We acknowledge Kostas Vilkelis, Isidora Araya-Day, Anton Akhmerov, and Jose Lado for fruitful discussions.
We also thank Valla Fatemi, Josep Ingla-Aynés, and Luca Banszerus for their input on the experimental implementation of our proposal.
We thank Anton Akhmerov, Josep Ingla-Aynés, and Kushagra Aggarwal for the feedback on the manuscript.


\section*{Data availablity}

The code and data generated for this manuscript and additional datasets are fully available on Zenodo~\cite{rigotti_manesco_2024_13284475}.

\bibliography{biblio.bib}

\newpage

\appendix

\section{Model and simulation details}
\label{sec:simulation_details}

All the numerical simulations presented in this manuscript were obtained via a tight-binding implementation of Bernal-stacked bilayer graphene.
The tight-binding model was implemented in Kwant~\cite{Groth2014}, following the Slonczewski-Weiss-McClure parametrization~\cite{PhysRev.108.612,PhysRev.109.272,PhysRev.119.606}
and using tight-binding parameters obtained via infrared spectroscopy~\cite{PhysRevB.80.165406}.
We neglect the effects of spin-orbit coupling and therefore treat spins as a trivial degeneracy.
The numerical implementation of the pristine Slonczewski-Weiss-McClure tight-binding Hamiltonian $\mathcal{H}_{\text{pristine}}$ follows exactly the procedure described in Ref.~\cite{Torres_Luna_2025} and implemented numerically in Ref.~\cite{zenodo}.
Throughout the manuscript, we simulate mesoscopic devices with sizes $\sim 1 \unit{\micro\meter}$.
To minimize the cost of the simulations, we rescale the tight-binding model by increasing the lattice constant as $\tilde{a} \mapsto s a$ and then adjusting the hopping parameters to preserve the low-energy Hamiltonian.
We set $s=10$ in the transport simulations and $s=20$ in the bandstructure simulations due to their larger computational cost.
To this end, we also follow the procedure developed in Ref.~\cite{Torres_Luna_2025}.
This reference also justifies the choices for $s$ for the parameter range used in the manuscript.

Besides the pristine Slonczewski-Weiss-McClure tight-binding Hamiltonian $\mathcal{H}_{\text{pristine}}$, we add onsite modulations of the chemical potential $\mu$ and sublattice imbalance $U$ to emulate the effects of multiple gate layers as
\begin{align}
 \mathcal{H}_{\text{onsite}} = \sum_{n} \psi_n^{\dagger}\left[U (\mathbf{r}_n) \mathrm{sign}(\mathbf{r}_n\cdot \hat{z}) - \mu(\mathbf{r}_n)  \right]\psi_{n},
    \label{eq:hamiltonian}
\end{align}
where $\psi_n = (c_{n}, c_{n})^T$,  $c_{n}$ is an annihilation operator of an electron at the atomic site $n$.
The screening effects are included by smoothening the electrostatic potential.
In our calculations, we set the potential at the interface between two regions with potentials $V_1$ and $V_2$ as
\begin{equation}
 V(\mathbf{r}) = \frac{(V_1 - V_2)}{2} \left[1 + \tanh\left(\frac{|\mathbf{r} - \mathbf{r}_B|}{\chi}\right)\right] + V_2~,
\end{equation}
where $\chi$ is the screening length, and $\mathbf{r}_B$ is the closest point at the boundary between the two regions.
Tipically $\chi\sim 25 - 50 \unit{\nano\meter}$ and we use $\chi=25 \unit{\nano\meter}$ thorughout the manuscript except if it is explicitly stated~\cite{Flr2022,Li2020_2}.
Thus, the chemical potential is the average between the potential in the two layers
\begin{align}
    \mu(\mathbf{r}) = \frac{V_{\text{upper}}(\mathbf{r}) + V_{\text{lower}}(\mathbf{r})}{2}~,
\end{align}
and the layer imbalance is the difference between the layer potentials
\begin{align}
 U(\mathbf{r}) = V_{\text{upper}}(\mathbf{r}) - V_{\text{lower}}(\mathbf{r})~.
\end{align}
We fix $U(\mathbf{r}) = \pm 100 \unit{meV}$ far from the gate edges through the manuscript.

We compute the conductance across the devices via scattering formalism.
All the calculations are performed at zero bias.
For the two-terminal devices simulated in the manuscript, the zero-bias conductance is
\begin{align}
 G = \frac{2e^2}{h} \mathrm{Tr} (tt^{\dagger})~,
\end{align}
where $t$ is the transmission matrix, and the factor of 2 comes from spin degeneracy.

\section{Gate-defined valley-helical channels}
\label{sec:displacement}

\paragraph{Graphene multilayers have an eletrically-controllable band gap.}

\begin{figure}[!t]
    \centering
    \includegraphics[width=\textwidth]{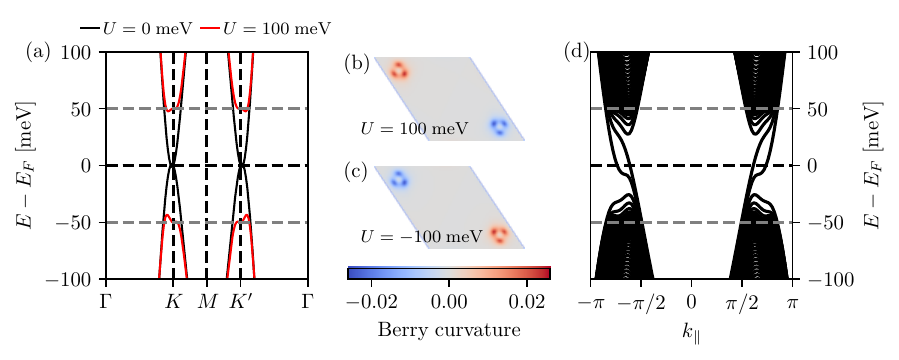}
    \caption{
        (a) Bulk bandstructure of Bernal bilayer graphene with $U_0=0$ and $U_0=100\unit{meV}$.
        A sublattice imbalance $U_0$ opens a gap in the band structure.
        Berry curvature for (b) $U_0=100 \unit{meV}$ and (c) $U_0=-100 \unit{meV}$.
        Due to time-reversal symmetry, the two valleys have opposite Berry curvature.
        However, the Berry curvature changes sign with layer imbalance.
        The valley Chern number changes across an interface where $U$ flips sign results in the in-gap valley helical modes shown in panel (d).
    }
    \label{fig:valley_berry}
\end{figure}

Due to a combination of mirror and time-reversal symmetries, the conduction and valence bands in graphene are degenerate at the corners of the Brillouin zone, as shown in Fig.~\ref{fig:valley_berry} (a).
This degeneracy makes graphene a gapless semiconductor.
Thus, modulations of the carrier density are insufficient to confine electrons due to Klein tunneling~\cite{Katsnelson2006,KATSNELSON200720}.
However, in graphene multilayers, the mirror symmetry is broken by the application of an out-of-plane electric field.
The potential imbalance $U$ between the layers opens a gap in the electronic structure~\cite{McCann_2006,PhysRevLett.99.216802,Min_2007,Oostinga2007,Zhang2009,PhysRevB.80.165406,Mak2009} as depicted by Fig.~\ref{fig:valley_berry} (a). 
Therefore, it is possible to control the band gap of graphene multilayers with double-gated devices~\cite{Mak2009,Weitz2010}.

\paragraph{Topological edge states propagate at the interface between regions with opposite displacement field.}

In Bernal bilayer graphene, the Berry curvature at each valley depends on the direction of the external electric field, shown in Fig.~\ref{fig:valley_berry} (b, c).
Thus, at the interface of two regions with opposite displacement fields, the valley Chern number changes.
As a consequence, these interfaces host topological valley helical channels.
In Fig.~\ref{fig:valley_berry} (d) we show the dispersion of a nanoribbon with a switching layer imbalance.
Due to the valley Chern number switch at this interface, in-gap valley-helical states propagate along the interface~\cite{Martin2008,Jung2011, Li2016}.
The helicity of these channels thus constrains the electronic motion in a single direction.
For this reason, these channels were proposed as valley filters.

\section{Valley filtering with split gates}
\label{app:split_gates}

\begin{figure}[t!]
    \centering
    \includegraphics[width=0.95\textwidth]{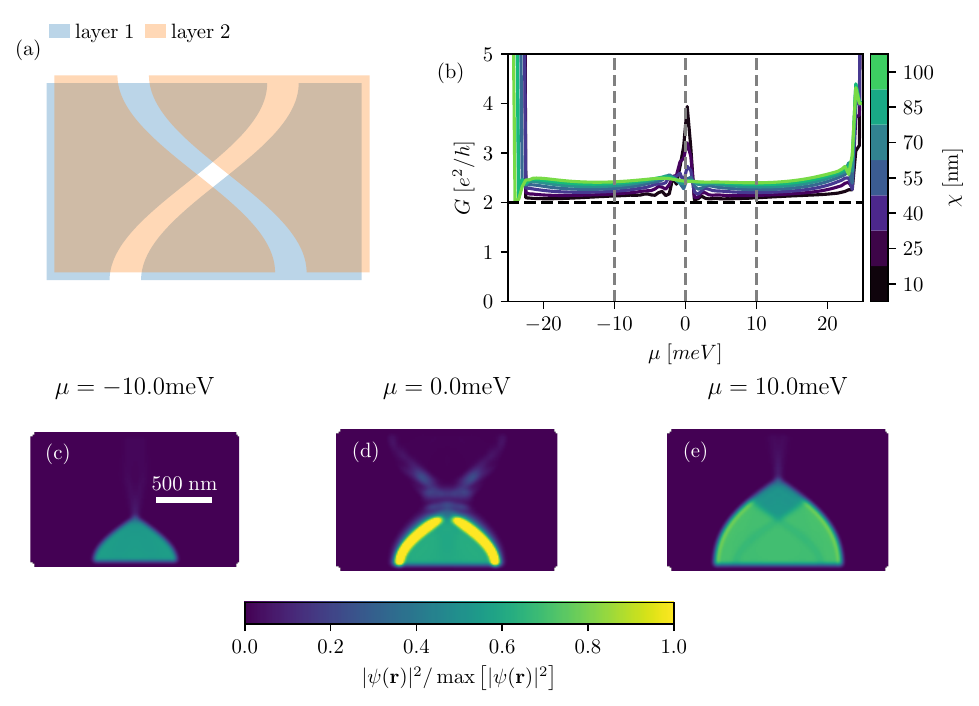}
    \caption{
        (a) Illustration of the gate layout with a single crossing point with split gates.
        (b) Conductance across a device with the layout geometry from panel (a) as a function of the global chemical potential $\mu$ and screening length $\chi$.
        We observe that the residual conductance above $4e^2/h$ increases with $\chi$ due to higher tunneling probabilities.
        The conductance shows a peak at $\mu=0$ due to additional transmission probabilities through the gap between split gates as we can observe by comparing the scattering wavefunctions in panel (d) with (c) and (e).
    }
    \label{fig:split_gates}
\end{figure}

\paragraph{The device also works with split gates.}

Due to possible complications in fabricating the layout in Fig.~\ref{fig:device} (a) with four layers of gates, we also calculate the conductance of a device with similar zigzag geometry but two layers of split shown in Fig~\ref{fig:split_gates} (a).
Despite the distance of $200\unit{\nano\meter}$ between gates used in the simulations, the transmission of non-topological in-gap states is still suppressed, as shown in Fig~\ref{fig:split_gates} (b).
Additionally to the conductance plateau at $G\approx 4e^2/h$, we also observe a conductance peak at charge neutrality.
By inspecting the scattering wavefunctions, shown in Fig.~\ref{fig:split_gates} (c-e), we conclude that this resonance peak occurs due to additional propagating modes within the split-gates gap.
Thus, we demonstrate that both the four-gate layout shown in Fig.~\ref{fig:device} (e) and the double-split-gate geometry in Fig.~\ref{fig:split_gates} (a) can be used to fabricate valley-helical quantum point contacts without gate-defined narrow channels.
We conjecture that the lack of conductance quantization in recent experiments~\cite{davydov2024easy} is due to the presence of these additional channels.
Therefore, we suggest that an additional gate layer can be used to reach quantized conductance, as recently done in geometries with narrow channels~\cite{Huang2024}.
However, verifying this conjecture requires a more realistic simulation of the electrostatic environment, which is beyond the scope of this work.

\section{Lattice implementation}
\label{sec:tangent}

Here we present a lattice model that describes the low-energy behavior of the superlattices described throughout the text.
We perform a lattice regularization of the Hamiltonian in Eq.~\ref{eq:h_eff} without fermion doubling following the tangent fermion approach~\cite{PhysRevD.26.468,Beenakker_2023}.
The core idea of this approach is obtaining a tangent dispersion for the helical states that can be expressed in a lattice as a generalized eigenproblem $H\Psi = ES\Psi$.
Particularly, for the Hamiltonian in Eq.~\ref{eq:h_eff}, we write for the helical
\begin{equation}
    H_{\mathrm{helical}} = i \sum_{i, \tau, \rho}\tau \left[\delta|i, \tau, \rho\rangle_h \langle i, \tau, \rho|_h + (t + \rho \delta t) |i + 1, \tau, \rho\rangle_h \langle i, \tau, \rho|_h\right]
\end{equation}
with $t = (v^+ + v^-) / 2a$, $\delta t = (v^+ - v^-) / 2a$.
The dot Hamiltonian follows trivially,
\begin{equation}
    H_{\mathrm{dot}} = \epsilon \sum_{i, \tau} |i, \tau \rangle_d \langle i, \tau|_d
\end{equation}
and the coupling between helical states is
\begin{equation}
    H_{\mathrm{\Delta}} = \Delta \sum_{i, \tau, \rho} |i, \tau, \rho\rangle_h \langle i, \tau|_d + h.c.~.
\end{equation}
We used the subindices $d$ and $h$ to destinguish states in the dot chain and helical lattice.
The overlap matrix of dot states corresponds to an orthogonal basis
\begin{equation}
    S_{\mathrm{dot}} = \sum_{i, \tau} |i, \tau\rangle_d \langle i, \tau|_d~.
\end{equation}
whereas for the helical modes
\begin{equation}
    S_{\mathrm{helical}} = \sum_{i, \tau, \rho}|i + 1, \tau, \rho\rangle_h \langle i, \tau, \rho|_h
\end{equation}
Solving this generalized eigenproblem with periodic boundary conditions result in the spectrum shown in Fig.~\ref{fig:helicity} (c).
While the approach results in a single Dirac cone at $k=0$, the tangent dispersion has a cusp at $k=\pm\pi$, and for this reason we plot the dispersion in Fig.~\ref{fig:helicity} (c) within $[-\pi / 2, \pi /2]$.

\end{document}